\documentstyle[twocolumn,aps]{revtex}

\newcommand{\simle}
{\raisebox{-0.75ex}[-1.5ex]{$\;\stackrel{<}{\sim}\;$}}
\newcommand{\simge}
{\raisebox{-0.75ex}[-1.5ex]{$\;\stackrel{>}{\sim}\;$}}
\def\e{{\epsilon}}
\def\k{{ {\bf k} }}

\def\q{{ {\bf q} }}

\def\w{{\omega}}
\def\a{{\alpha}}

\def\l{{\lambda}}

\begin{document}
    \sloppy


\draft
\title{
Phase Diagram of Superconductivity \\
on the Anisotropic Triangular Lattice Hubbard Model
}

\author{
Hiori {\sc Kino}
and 
Hiroshi {\sc Kontani}$^{1}$
}

\address{
Joint Research Center for Atom Technology, Tsukuba,
Ibaraki, 305-8562, Japan.
\\
$^1$Institute for Solid State Physics, University of Tokyo,
7-22-1 Roppongi, Minato-ku, Tokyo 106-8666, Japan.
}

\date{\today}

\maketitle

\begin{abstract}
We study the electronic states of the anisotropic triangular lattice
Hubbard model at half filling, which is a simple effective 
model for the organic superconducting $\kappa$-BEDT-TTF compounds.
We treat the effect of the Coulomb interaction 
by the fluctuation exchange (FLEX)  method, and obtain 
the phase diagram of this model for various sets of parameters.
It is shown that
the $d$-wave superconductivity is realized in the wide region 
of the phase diagram, next to the antiferromagnetic states.
The obtained phase diagram explains the characters of the experimental
results very well.
\end{abstract}


It is known that the layered organic compounds
made of BEDT-TTF molecules exhibit rich variety of ground states,
through the strong correlation effects  between electrons.
Among them, $\kappa$-polytype of BEDT-TTF systems
has attracted much attention because many of them are superconducting 
materials. The superconducting transition temperature ($T_c$) 
reaches $\sim 10$K, 
whose ratio to the band width is as large as
that of the high-$T_{\rm c}$\ cuprates.

In fig. 1, we show the schematic pressure-temperature phase diagram of 
the $\kappa$-(BEDT-TTF) compounds.\cite{Kanoda-review}
For example, 
Cu[N(CN)$_2$]Cl salt at ambient pressure
 is 
in the antiferromagnetic (AF) insulating phase 
 as shown by the arrow in the figure.
The N\'eel temperature ($T_{\rm N}$) is 27K and the spin moment is 
greater than 0.4$\mu_{\rm B}$\ per magnetic unit, i.e., 
dimer of BEDT-TTF molecules.
The antiferromagnetism disappears with increasing pressure
to give birth to the superconductivity.
$T_{\rm c}$ takes its maximum value, 13K, at 200bar,
but this superconducting (SC) phase also suppressed by several kbar,
where a paramagnetic metallic phase is stabilized.
This means that the superconductivity coexists with 
the strong antiferromagnetic correlations.
Thus, it is natural to consider that the antiferromagnetic fluctuation 
is the origin of superconductivity.

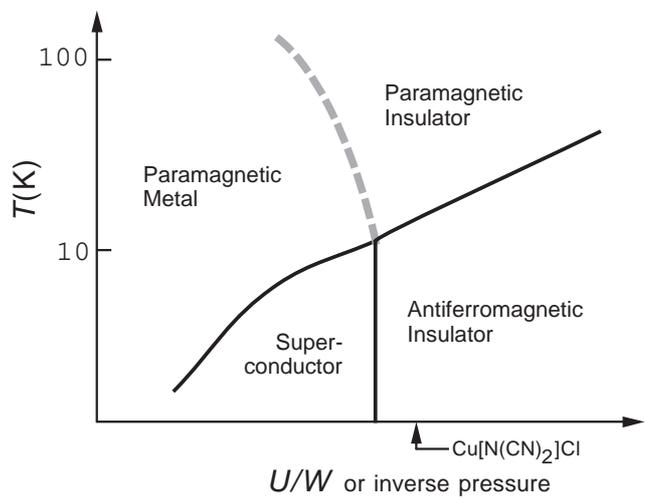
\begin{figure}
\caption{Schematic phase diagram of $\kappa$-(BEDT-TTF) compounds.}
\label{schematic-diagram}
\end{figure}

In the previous studies, the organic compound has been studied
theoretically through the local density approximation 
and the Hartree-Fock approximation.
 \cite{LDA,HF-approx}
The latter study could explain the pressure dependence of 
the AF boundary in fig. \ref{schematic-diagram}.
The structures of these compounds are very complex.
Nonetheless, the electronic structure near the Fermi level is very simple,
because it is composed of well separated HOMO levels.
The simplest effective model is a triangular 
lattice Hubbard model with anisotropic hoppings 
at half filling, as shown below.
The effective on-site Coulomb repulsion ($U$) is 
estimated to be comparable with the band width ($W$) in this model.

The aim of this letter is to discuss the mechanism of the
superconductivity, and to understand the entire phase diagram shown
in fig. 1.
For this purpose, we study the anisotropic triangular lattice
Hubbard model by using the fluctuation exchange (FLEX) method.
It is one of the methods that have much advantage
for the systems with large spin fluctuations
and that can handle the SC and AF states on the same footing.
In this letter, we find the $d$-wave SC phase next to the AF phase.
The obtained $T_c$ is consistent with experiments.
This is the first systematic study of the organic SC compounds
by using the FLEX method.

The FLEX method is a kind of self-consistent perturbation theory 
with respect to $U$, which gives reliable results although
it is an approximation.
For example, the imaginary time Green function obtained by
the FLEX method agrees well with that 
by the QMC for the square lattice Hubbard model with moderate $U$.
 \cite{FLEX_DMRG}
This method has been  applied to the study of high-$T_{\rm c}$,
and various non-Fermi liquid behaviors are reproduced well.
 \cite{high-Tc-Monthoux,2D-SC-Monthoux,high-Tc-Dahm,shadow-band,incipient-AF,d-p-model,d-p-model2}
Recently it has also been  applied to the superconducting ladder compound, 
Sr$_{14-x}$Ca$_x$Cu$_{24}$O$_{41}$.
 \cite{Trellis}

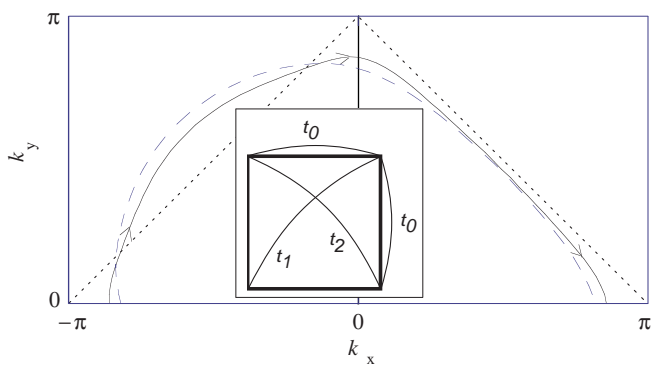
\begin{figure}
\caption{A model and the Fermi surface when $t_1=-0.5$\ at $T$=0.024. The solid line shows the one at $U$=4.6 and the dotted line, $U$=0 . Arrows on the solid line are explained in fig.4.
64$\times$64 $k$-points and 512 Matsubara frequencies are used.
}
\label{model_and_fs}
\end{figure}

Dimerization of $\kappa$-(BEDT-TTF) salts is so large
that we can take a dimer of BEDT-TTF molecules as a unit. 
So, as a simple effective model, 
we consider the two-dimensional single band Hubbard model 
with the lattice structure shown in fig.\ref{model_and_fs},
where  $t_0$, $t_1$, $t_2$  are  hopping parameters.
We also take into account the on-site Coulomb repulsion ($U$),
which is an energy loss when two electrons are on the same BEDT-TTF dimer
in the original $\kappa$-(BEDT-TTF) lattice.
The hopping parameters depend on materials, but
both the band calculations and experiments
indicate that 
$t_1\sim t_0$\  and $|t_2| \ll |t_0|,|t_1|$.
 \cite{LDA,CN3-singleband-kappa,NCS-SdH,NCNCl-SdH}
In this study, we put $t_0=-1$, $t_2=-0.05$, and 
change $t_1$(<0) as a parameter.
In this sense, this model is an anisotropic triangular lattice.
For $|t_1|\le|t_0|$, 
$W$\ is always 8$|t_0|$.
A similar model has been studied recently, 
but the essence of the physics of $\kappa$-(BEDT-TTF) salts
is contained  in the above simple model.\cite{Schmalian}

Here, we explain the FLEX method.
The Dyson equation is written as
\begin{eqnarray}
\left\{ {G}(\k,\e_n) \right\}^{-1}
= \left\{ {G}^0(\k,\e_n) \right\}^{-1} - {\Sigma}(\k,\e_n),
 \label{eqn:Dyson}
\end{eqnarray}
where ${G}^0(\k,\e)$ is the unperturbed Green function.
The self-energy is given by
\begin{eqnarray}
& &\Sigma(\k,\e_n) 
 = T\sum_{\q,l} G(\k-\q,\e_n-\w_l)\cdot U^2    \nonumber \\
& &\ \times \left( \frac32 {\chi}^{(-)}(\q,\w_l) 
  +\frac12 {\chi}^{(+)}(\q,\w_l) 
  - {\chi}^0(\q,\w_l) \right) \mbox{,}
     \label{eqn:self} \\
& &{\chi}^{(\pm)}(\q,\w_l)
 = {\chi}^0 \cdot \left\{ {\hat 1} \pm 
  U{\chi}^0(\q,\w_l) \right\}^{-1} \mbox{,} 
     \label{eqn:chi} \\
& &\chi^0(\q,\w_l)
 = -T\sum_{\k, n} G(\q+\k,\w_l+\e_n) G(\k,\e_n) \mbox{,}
     \label{eqn:chi0}
\end{eqnarray}
where $\e_n= (2n+1)\pi T$ and $\w_l= 2l\pi T$, respectively.
We solve the equations (\ref{eqn:Dyson})-(\ref{eqn:chi0}) self-consistently,
choosing the chemical potential $\mu$ 
so as to keep the system at half-filling.

To determine $T_N$, we calculate the Stoner factor 
without vertex corrections, $\a_{\rm S}$, given by
\begin{eqnarray}
\a_{\rm S}= \max_{\k}\left\{ \ U\cdot \chi^{0}(\k,\w\!=\!0)\  \right\},
 \label{eqn:Stoner}
\end{eqnarray}
The antiferromagnetic critical points are determined
by the Stoner criterion, $\a_{\rm S}= 1$.

We also determine $T_{\rm c}$
by solving the linearized Eliashberg equation with respect to 
the singlet-pairing order parameter,
$\phi(-\k,\e_n)= + \phi(\k,\e_n)$,
\begin{eqnarray}
& &\lambda \cdot \phi(\k,\e_n)= -T\sum_{\q, m}
 V(\k-\q,\e_n-\e_m)  \nonumber \\
& & \ \ \times G(\q,\e_m) G(-\q,-\e_m)
 \cdot \phi(\q,\e_m), \label{eqn:lambda}\\
& &{V}(\k,\w_l)=  \frac32 U^2 {\chi}^{(-)}(\k,\w_l)
 - \frac12 U^2 {\chi}^{(+)}(\k,\w_l) + U ,
 \label{eqn:V}
\end{eqnarray}
where $T_{\rm c}$ is given by the  condition that $\lambda=1$.

The theories  by Mermin, Wagner and Hohenberg
 prohibits finite  $T_{\rm N}$ and $T_{\rm c}$\ 
in two dimension.
 \cite{Mermin-Wagner,Hohenberg}
It is well known that $\a_{\rm S}$ given by  eq.(\ref{eqn:Stoner}) satisfies
this condition, 
because FLEX treats the spin fluctuations self-consistently.
 \cite{high-Tc-Dahm,incipient-AF}
So we determine $T_{\rm N}$\ by the condition $\a_{\rm S}= \a_{\rm N}$,
where we set $\a_{\rm N}$ as $(1-\a_{\rm N})^{-1} \sim O(100)$.
The AF state will occur through the weak coupling between layers.
On the other hand, $\l=1$ is fulfilled at  finite $T_{\rm c}$\ 
by using eq.(\ref{eqn:lambda}).
The obtained $T_{\rm c}$\ is, however, reliable in many cases.



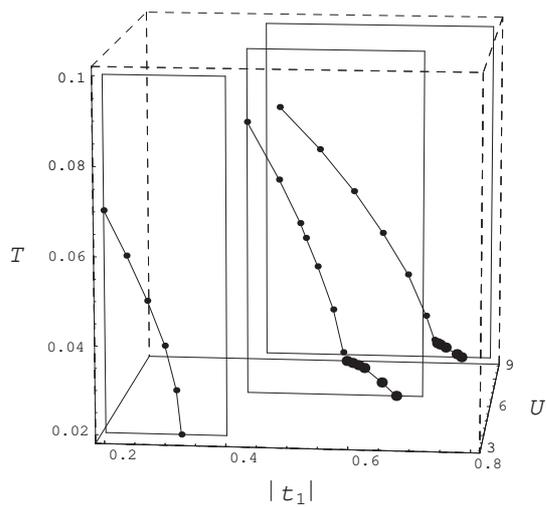
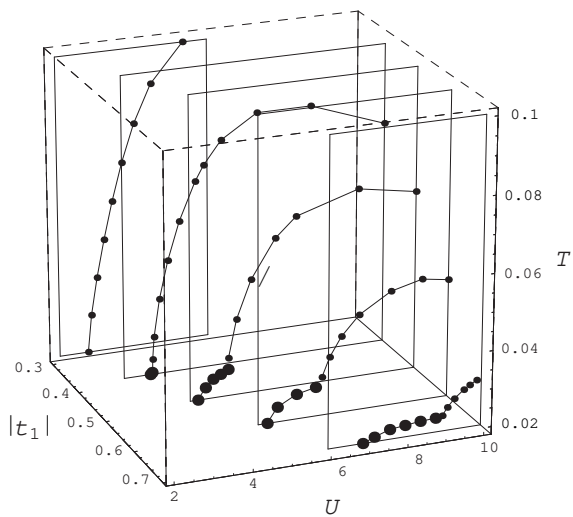
\begin{figure}
\caption{Calculated phase diagram on the planes of 
 $U$\ vs $T$\ at several values of  $|t_1|$\ (a), 
and on the planes of $|t_1|$\ and $T$\ at a few $U$\ (b), respectively.
The antiferromagnetism is realized in 
the temperature region below the small circles.
and the superconductivity, below the large circles.
Lines are guides for the eye.
}
\label{U_t_T_diagram}
\end{figure}

First, let us see the phase diagram 
as a function of $|t_1|(=-t_1)$, $U$\ and $T$, for $T\ge 0.02$.
A number of sections of the phase diagram are plotted 
in fig.\ref{U_t_T_diagram}.
16$\times$16 $k$-points  and 256 Matsubara frequencies 
(abbreviated as 16$\times$16$\times$128 mesh)
are used to draw this phase diagram.
For 32$\times$32$\times$1024 or  64$\times$64$\times$512 mesh
 the Stoner factor becomes larger,
 but that effect can be compensated 
by taking slightly larger $\alpha_{\rm N}$.
In fig.\ref{U_t_T_diagram} we set  $\alpha_{\rm N}=0.99$.
As for the superconductivity, the size effect of $T_{\rm c}$\ is
rather smaller than that of  $T_{\rm N}$,
because 
$\lambda$\  rapidly grows  near the critical temperature.

Fig.\ref{U_t_T_diagram}(a) shows the critical temperatures, 
$T_{\rm N}$ (small circles)  and  $T_{\rm c}$ (large circles),
 on the planes of $U$\ vs $T$\ at several values of  $|t_1|(=-t_1)$.
Only AF phase appears at $|t_1|$=0.3, 
 SC occurs for $|t_1|=0.4\sim0.7$\ on the left hand side of 
 the AF phases.
As $|t_1|$ \ increases, we  see the followings:
The onset value of  $U$\ for the  AF region  
 shifts larger,
at the same time, the peak of $T_{\rm N}$\ decreases, 
and the SC region are extended.
The maximum value of $T_{\rm c}$ \ at each section
increases for $|t_1|<0.6$, 
and then decreases for  $|t_1|>0.6$. 

 Fig.\ref{U_t_T_diagram}(b) shows  $T_{\rm c}$\ and $T_{\rm N}$\ 
on the planes of $|t_1|$\ vs $T$,
 at $U$=3.0, 6.0 and 9.0 for $T\ge 0.02$.
At  $U=3.0$ 
only the AF phase appears.
On the other hand, 
the SC state  occurs  next to the AF phases at $U$=6.0 and 9.0.
As $U$\ increases, both the SC and AF boundaries move 
toward larger $|t_1|$\ region.
At $U$=9.0, which is the largest $U$\ in this figure, 
$T_{\rm N}$\  decreases as $|t_1|$\ becomes larger 
and the SC state takes place instead of  the AF state
for $|t_1|>$0.7.
This SC phase , 
however, disappears for $|t_1|>$0.75.
At the $|t_1|$=0.8, the AF occurs at larger $U$, 
but 
 no SC state is stabilized at $T\ge 0.02$.
No SC and AF phases appear at $|t_1|$=0.9 and 1.0
for $U\le 40$.  
We will discuss the reason for this in the latter section.
The value of $\lambda$
 increases monotonically  not only as $T$ is decreased.
but also as decreasing $|t_1|$\ 
as shown in fig.\ref{U_t_T_diagram}

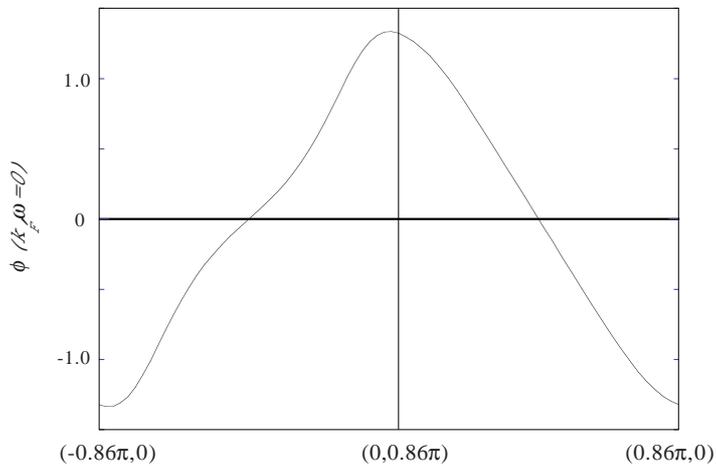
\begin{figure}
\caption{The momentum dependence of $\phi(k_F,\omega=0)$\
along the line with the arrows on the Fermi surface in fig.2}
\label{wf_SC}
\end{figure}

Next in fig.\ref{wf_SC},
we plot
the SC order parameter, 
$\phi(k,\omega=0)$,
along the line with the arrows on  the Fermi surface.
The parameters used in fig.\ref{wf_SC} are
 $U$=4.6, $t_1$=$-0.5$ and $T=0.024$,
 where $\lambda=1.02$.
One can clearly see that there are nodes in the  ($\pi$,$\pi$) and
($-\pi$,$\pi$) directions 
and the superconductivity has
$d_{x^2-y^2}$-like symmetry.
The maximum values of $\phi(k_F,\omega=0)$\ is slightly shifted from
$(0,\pi)$\ direction to $(-\pi,\pi)$\ direction.
This is also the case for all the other superconducting phases.
We have checked 
that 
the odd-parity superconductivity is not realized
in the present model.

%

To understand the overall aspects of the phase diagram,
we discuss it on the basis of the weak coupling theory.
In this  scheme, the effective interaction between a singlet Cooper pair
is given by
\begin{eqnarray}
& &\lambda_{\rm w}= - \max_{\phi}
 \langle V(q-q',0) \phi(q)\phi(q')\rangle_{\rm FS}/
\langle|\phi(q)|^2\rangle_{\rm FS},
\end{eqnarray}
where $\phi(q)$ is the energy-independent SC order parameter,
and $V(q,\w)$ is given by eq.(\ref{eqn:V}).
The obtained $\phi(q)$ has the d-wave-like  symmetry in this model
as shown above.
Introducing the cutoff energy $\w_c$, the $T_c$ within the weak
coupling theory is given by 
$ T_c= 1.14\w_c \exp(-1/\lambda_{\rm w})$.

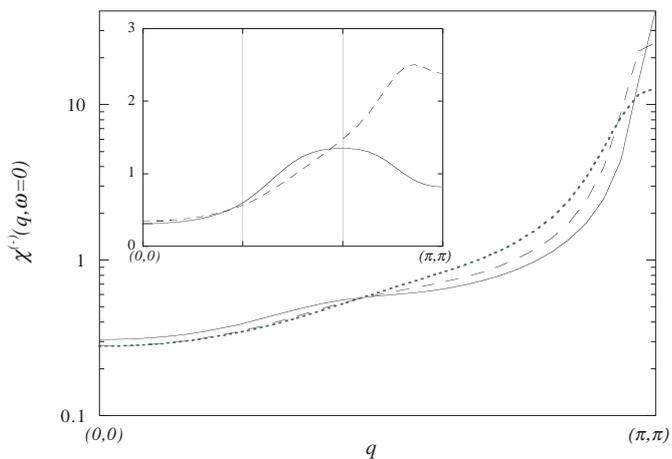
\begin{figure}
\caption{
$\chi^{(-)}(q,\omega)$\ from $q$=(0,0)  to ($\pi$,$\pi$)
at
$t_1=-0.3$, $T=0.03$ and $U=2.85$ (solid line),
$t_1=-0.5$, $T=0.03$ and $U=5.00$ (broken line) and
$t_1=-0.7$, $T=0.03$ and $U=9.60$ (dotted line).
$\max_k U \chi^{0}(k,\omega=0)$\ is around 0.992
at $k$=($\pi$,$\pi$) for all the cases.
The inset shows $\chi^{(-)}(q,\omega)$\ from $q$=(0,0)  to ($\pi$,$\pi$)
at
$t_1=-1.0$, $t_2=0$, $T=0.025$ and $U=15$ (solid line)
and at $t_1=-1.0$, $t_2=0.05$, $T=0.02$ and $U=15$ (broken line).
$\max_q U \chi^{0}(q,\omega=0)$\ at $t_1=-1.0$\ is about 0.953
at $q$=($2\pi/3$,$2\pi/3$).
The 64$\times$64$\times$512 meshes are used.
}
\label{susc}
\end{figure}

In fig.\ref{susc},
we show the obtained $\chi^{(-)}(q,\w=0)$ by the FLEX method
for $|t_1|=0.3\sim0.7$. 
At each $|t_1|$, we choose $U$ so as to satisfy $\a_{S}= 0.99$.
Then, $V(q) \approx 3U^2/2 \cdot \chi^{(-)}(q,0)$
and $\max_q\{V(q)\} \approx 3U/2(1-\a_{S})$.
For each $|t_1|$, $\chi^{(-)}(q,0)$ shows a sharp peak around
$(\pi,\pi)$, which  favors the $d$-wave SC state
similarly to the high-$T_{\rm c}$ cuprates.
We note that $\chi^{(-)}(q,0)$ has an incommensurate 
structure when $\a_{S}$ is rather small.

As $|t_1|$ increases from 0.3 to 0.7, 
the peak of $\chi^{(-)}(q,0)$ around $(\pi,\pi)$ becomes broader,
because the nesting condition becomes worse.
This situation will make $T_{\rm c}$ lower within the weak coupling scheme.
On the other hand, as $|t_1|$ increases, the value of $\max_q\{V(q)\}$
becomes larger, which will make $T_{\rm c}$ higher.
The optimum $|t_1|$ for SC, which is $|t_1|=0.6$ in our calculation,
may be  determined by the balance of these two opposite  effects.
We have already shown that $d$-wave SC occurs at $T\ge0.02$ 
in the case of $|t_1|=0.4\sim0.7$.
We also find $T_{\rm c} \approx 0.01$ at $|t_1|=0.8$.

At $|t_1|=0.3$, AF phase completely prevails over the SC phase because 
$\max_q\{V(q)\}$ is rather small at $\a_{S}= 0.99$.
On the other hand, when $|t_1| \simge 0.9$, $\a_{S}= 0.99$ \ 
is not satisfied for $U<40$.
So, $\max_q\{V(q)\}$ is also small.
Besides, the structure of $V(q)$ changes drastically 
as shown in the inset of fig.\ref{susc}.:
The peak position of $\chi^{(-)}(q,0)$ shifts from 
$(\pi,\pi)$ towards $(2\pi/3,2\pi/3)$, so the AF correlation is violated.
If we map our model to the corresponding Heisenberg spin system on the 
anisotropic triangular lattice, the ground state classically changes 
from AF order to the 120
degree structure at $J_1/J_0 = (t_1/t_0)^2 = 2/3$, i.e., $t_1/t_0 = 0.82$.
The momentum $(2\pi/3,2\pi/3)$ corresponds to the 120 degree structure.
As a result, for $|t_1|\simge0.9$, any singlet SC state is not realized 
up to $U\le 40$.



The  Fermi surfaces at $t_1=-0.5$\ and $T=0.024$
 are drawn in fig.\ref{model_and_fs}.
The change of the Fermi surface is attributed to the renormalization
of $|t_1|$\ by many body effect.
The sum rule, which must be satisfied in principle,
 but not necessarily fulfilled in actual
numerical calculations, is checked to be satisfied within 0.5\%.
Let us discuss the connection between the real materials and 
the obtained phase diagram.
The $T_{\rm c}$\ of $\kappa$-(BEDT-TTF) is as high as 10K.
If one takes the noninteracting band width ($W$) to be 5000K, then 
0.03$|t_0|$=18K, which corresponds well with the experiments.
\cite{Kanoda-review}


Next we consider the effect of pressure.
First we return to the original $\kappa$-BEDT-TTF crystal.
It is natural to assume that 
the effect of pressure is most sensitive to  the loosely-packed parts,
i.e., to the interdimer ones.
The interdimer transfer integrals determine the band structure.
So  $t$'s or $W$ in the present model is sensitive to the external pressure,
while the interdimer transfer integrals and the on-site Coulomb energy 
on the BEDT-TTF molecule, or $U$ in the effective model, 
is insensitive to the applied pressure.
The ratio of $t_1/t_0$\ will change, 
but both $t_0$ and $t_1$ are estimated from the interdmer transfer integrals, 
thus the ratio does not change so large as $U/W$.
Because $W$\ is fixed in fig.\ref{U_t_T_diagram},
one can regard  the pressure effect as decreasing $U$.

Some band calculations suggest that $|t_1|\sim |t_0|$,
where no SC and AF phases appear in fig.\ref{U_t_T_diagram}, 
however.\cite{CN3-singleband-kappa}
One possibility for this discrepancy is that the band calculations estimate 
the value of $t_1/t_0$ larger.
The other is that the current  model is too simple to reflect multiband effects
properly.
Such a effect is probably taken into account 
by the renormalization of $t_1/t_0$ 
of this effective model, 
because 
 the  AF phase will emerges at $J_1/J_0\simle2/3$, or $t_1/t_0\simle0.82$\ 
in the present model as noted above.
If one choose such a $t_1/t_0$, 
then one can find not only the AF state, but also 
the SC state next to the AF phase. 
Moreover the SC phase appears in  higher pressure region.
In this sense, the phase diagram in fig.\ref{U_t_T_diagram} agrees well
with the experimental phase diagram.

%
Furthermore, 
the experimental AF state is commensurate,
 \cite{Kanoda-review}
which is also consistent with the FLEX results that
the peak of $\chi(q,0)$\ is always at ($\pi$,$\pi$) in the AF phase, 
 due to the self-consistency. 
Although the system is half-filling, the region where SC state appears 
in fig.\ref{U_t_T_diagram}
will not be replaced by the nonmagnetic Mott insulating one,
 because $U$ is smaller than $W$\ in that region.
We finally make a comment for chemistries that
$T_{\rm c}$ takes the maximum value, $\sim 0.029$, at $t_1\sim 0.6$.
$T_{\rm c}$ possibly becomes higher than that of the currently known materials
if one can compose  such compounds,
though the correspondency between the present simple model and 
the original complex structure 
is not straightforward. 

%

In conclusion, 
we have studied the anisotropic triangular lattice Hubbard models at half-filling  by using the FLEX method.
We obtained the phase diagram,
where  the $d$-wave superconducting phase  is realized next to the antiferromagnetic one.
The resultant phase diagram  is consistent with the experimental phase diagram in $\kappa$-(BEDT-TTF) systems.
These results suggest that  the electron-electron correlation is important 
in $\kappa$-(BEDT-TTF) compounds
 and that the spin fluctuation mechanism is the possible origin 
of the superconductivity.

We are grateful to K. Terakura, K. Yamada, K. Ueda, K. Kanoda, 
K. Yonemitsu, H. Kohno, T. Komatsu and M. Mori
 for valuable comments and useful discussions and encouragements.
This work is partly supported by NEDO.


\hrule

\begin{thebibliography}{99}
%
\bibitem{Kanoda-review} {\it A recent compact review is} 
K. Kanoda, Physica C {\bf 282-287}  299, (1997).
%
\bibitem{LDA} Y. -N Xu, W. Y. Ching, Y. C. Jean and Y. Lou,
Phys. Rev. B {\bf 52} 12946,  (1995).
%
\bibitem{HF-approx} H. Kino and H. Fukuyama,
J. Phys. Soc. Jpn. {\bf 65}  2158, (1996).
%
\bibitem{FLEX_DMRG} N. E. Bickers, D. J. Scalapino and S. R. White, Phys. Rev. Lett. {\bf 62} 961, (1989).
%
\bibitem{high-Tc-Monthoux} P. Monthoux and D. Pines, 
Phys. Rev. B {\bf 47} 6069, (1993).
%
\bibitem{2D-SC-Monthoux} P. Monthoux and D. J. Scalapino, 
 Phys. Rev. Lett. {\bf 72}  1874,  (1994).
%
\bibitem{high-Tc-Dahm} T. Dahm and L. Tewordt, 
Phys. Rev.  B {\bf  52} 1297, (1995).
%
\bibitem{shadow-band} M. Langer, J. Schmalian, S. Grabowski, 
and K. H. Bennemann, Phys. Rev. Lett. {\bf 75} 4508,  (1995).
%
\bibitem{incipient-AF} J. J. Deisz, D. W. Hess and J. W. Serene, 
Phys. Rev. Lett. {\bf 76}  1312, (1996).
%
\bibitem{d-p-model} S. Koikegami, S. Fujimoto and K. Yamada, 
J. Phys. Soc. Jpn. {\bf 66} 1438,  (1997).
%
\bibitem{d-p-model2} T. Takimoto and T. Moriya,
J. Phys. Soc. Jpn. {\bf 66} 2459,  (1997).
%
\bibitem{Trellis} H. Kontani and K. Ueda,  Phys. Rev. Lett. {\bf 80} 5619 (1998).
%
\bibitem{CN3-singleband-kappa} e.g., T. Komatsu, N. Matsukawa, T. Inoue and G. Saito, J. Phys. Soc. Jpn. {\bf 65} 1340, (1996).
The overlap integrals are also listed 
 for a number of $\kappa$-(BEDT-TTF) salts and related materials.
%
\bibitem{NCS-SdH}  K. Oshima, T. Mori, H. Inokuchi, H. Urayama, 
H. Yamochi and G. Saito, Phys. Rev. B {\bf 38} 938, (1988).
%
\bibitem{NCNCl-SdH} Y. Yamauchi, M. V. Kartsovnik, T. Ishiguro, M. Kubota and G. Saito,  J. Phys. Soc. Jpn. {\bf 65} 354, (1996).
%
%
%
\bibitem{Schmalian} In finishing this letter, we noticed the preprint:
 J. Schmalian, cond-mat/9807042.
%
\bibitem{Mermin-Wagner} N. D. Mermin and H. Wagner, Phys. Rev. Lett. {\bf 17} 1133, (1966).
%
\bibitem{Hohenberg} P. C. Hohenberg, Phys. Rev. {\bf 158} 383, (1967).
%
\end{thebibliography}
\end{document}